\documentclass[conference,compsoc]{IEEEtran}

\usepackage{booktabs}
\usepackage{graphicx}
\usepackage{mathtools}
\usepackage{color}
\usepackage{soul}
\usepackage{url}
\usepackage{hyperref}
\usepackage{tabularx}
\usepackage{booktabs}
\usepackage{graphicx}
\usepackage{multirow}
\usepackage[table,xcdraw]{xcolor}
 \usepackage[normalem]{ulem}
 \useunder{\uline}{\ul}{}
\usepackage{caption}
\usepackage{longtable}
\usepackage{multirow}
\usepackage{epsfig,endnotes}
\usepackage[most]{tcolorbox}
\usepackage{xcolor}

\ifCLASSOPTIONcompsoc
  \usepackage[nocompress]{cite}
\else
  \usepackage{cite}
\fi
\ifCLASSINFOpdf
\else
\fi
\begin{document}
%
\title{SoK: Privacy Personalised - Mapping Personal Attributes \& Preferences of Privacy Mechanisms for Shoulder Surfing}



%

\author{\IEEEauthorblockN{Habiba Farzand\IEEEauthorrefmark{1},
Karola Marky\IEEEauthorrefmark{3}, and
Mohamed Khamis\IEEEauthorrefmark{2} 
\IEEEauthorblockA{\IEEEauthorrefmark{1}University of Bristol,
United Kingdom
\IEEEauthorblockA{\IEEEauthorrefmark{2}University of Glasgow,
United Kingdom
\IEEEauthorblockA{\IEEEauthorrefmark{3}Ruhr University Bochum,
Germany}}}}}


\maketitle

\begin{abstract}
Shoulder surfing is a byproduct of smartphone use that enables bystanders to access personal information (such as text and photos) by making screen observations without consent. To mitigate this, several protection mechanisms have been proposed to protect user privacy. However, the mechanisms that users prefer remain unexplored. This paper explores correlations between personal attributes and properties of shoulder surfing protection mechanisms. For this, we first conducted a structured literature review and identified ten protection mechanism categories against content-based shoulder surfing.  We then surveyed N=192 users and explored correlations between personal attributes and properties of shoulder surfing protection mechanisms. Our results show that users agreed that the presented mechanisms assisted in protecting their privacy, but they preferred non-digital alternatives. Among the mechanisms, participants mainly preferred an icon overlay mechanism followed by a tangible mechanism. We also found that users who prioritized out-of-device privacy and a high tendency to interact with technology favoured the personalisation of protection mechanisms. On the contrary, age and smartphone OS did not impact users' preference for perceived usefulness and personalisation of mechanisms. Based on the results, we present key takeaways to support the design of future protection mechanisms.  
\end{abstract}


%
\IEEEpeerreviewmaketitle

\section{Introduction}
By the end of 2023, 70\% of the world's population was using smartphones~\cite{smartphoneusersglobally}. Smartphone users are increasing every day and are expected to reach 62.53 million users alone in the UK by 2029~\cite{smartphoneusers}. Smartphones are no longer just phones but the interfaces to users' personal lives. With smartphones comes the threat of privacy invasions, with shoulder surfing being one of the most frequently reported privacy threats~\cite {farzand2022shoulder}. Shoulder surfing - belonging to the novice attack category in the categorisation of social engineering and side-channel attacks, refers to observing someone's device screen without permission~\cite{eiband2017understanding,farzand2024systematic}. Possibly, the reason behind the existence of shoulder surfing lies in its ease of execution. For example, shoulder surfing only requires an observer to be close to the victim user and carefully observe the device content~\cite{farzand2024systematic}. Shoulder surfed content can be classified into two categories: authentication, such as passwords or PINs~\cite{khamis2017gazetouchpin}, and content, such as photos or text~\cite{eiband2017understanding,farzand2022hate,farzand2022shoulder}. While both categories of shoulder surfing have been reported by users, content-based shoulder surfing is more frequently reported by users~\cite{farzand2022shoulder,eiband2017understanding,farzand2021interplay,farzand2022hate}. To protect users' privacy from content-based shoulder surfing, researchers have proposed several device-based mechanisms that range from applying a filter over the full screen, such as grayscale~\cite{zhou2016enhancing}, to customized content hiding, such as blackout or crystallise filter\cite{khamis2018eyespot,von2016you}.

While these mechanisms may deliver effectiveness in terms of security and usability, they may not be ideal for every user group as every user has their preferences and needs. Therefore, applying a one-size-fits-all approach to all users is challenging. This is also evident from research on user interface design that a one-size-fits-all solution rarely meets the demand for appreciable user experiences~\cite{cooper2014face}. This signifies the importance of designing mechanisms that tailor every user group's needs and preferences. Providing user experience while considering users' personal attributes advocates for users' tailored privacy. Further, the literature presents several examples that provide evidence that personal attributes such as age, gender, and technical affinity impact users' privacy-related behaviors~\cite{alsoubai2022permission,liu2014reconciling}. 

However, the present research lacks an understanding of how personal attributes shape a user's preference for protection mechanisms against shoulder surfing. Knowing how to build user-centred privacy protection mechanisms targeted at specific user groups is important. To address this missing puzzle piece, our first research question under investigation is: 

\vspace{-1mm}
\begin{description}
    \item[RQ$_1$:] \textbf{Protection Mechanisms: }What are the existing protection mechanisms against shoulder surfing?
\end{description}  
\vspace{-1mm}



To inform the design of protection mechanisms, the first step is to understand what properties of protection mechanisms are important to users. For example, in an interview study, users reported having a mechanism that communicates irrelevant information to the observers so that the observer knows that they have been caught shoulder surfing by the user~\cite{farzand2021interplay}. Therefore, we assess the properties of mechanisms: ability to alert the user, ability to mitigate shoulder surfing, conveying irrelevant information to the bystander, conveying bystander information, and unnoticeable (non-visual) mechanisms. This is important for future researchers who are focusing on the design of protection mechanisms. This motivates our next research question: 

\vspace{-1mm}
\begin{description}
    \item[RQ$_2$:] \textbf{Perception of Privacy Protection Mechanisms Against Shoulder Surfing: }How do users perceive different privacy protection mechanisms developed for protection against shoulder surfing?
\end{description}  
\vspace{-1mm}

The next step towards designing user-centred protection mechanisms is to look into how users' personal attributes correlate with their perceptions of specific properties of protection mechanisms. To this end, our final research question in this investigation is: 

\vspace{-1mm}
\begin{description}
    \item[RQ$_3$:] \textbf{Personal Attributes \& Properties for Protection Mechanisms: }Which personal attributes (for example, age, gender, importance for out-of-device privacy) correlate with preferences towards properties of protection mechanisms?
\end{description}  
\vspace{-1mm}

To answer the above-listed research questions, we opted for descriptive research that focuses on already existing protection mechanisms~\cite{stangor2014introduction,distler2021systematic}. We conducted a systematic literature review to answer \textbf{RQ$_1$} and narrowed down protection mechanisms against content-based shoulder surfing for smartphones. We then conducted an online survey with a sample from the UK (N=192) to answer \textbf{RQ$_2$,$_3$}. The questionnaire focused on exploring people's general perception of those protection mechanisms and capturing personal attributes that describe our participants, such as age, gender, importance for out-of-device privacy or affinity for technology.

We found that participants agreed that the presented mechanisms offered privacy protection. However, they preferred non-digital alternatives, such as covering the screen using their hands, to device-level protection mechanisms, such as screen visibility filters. Moreover, participants expressed they would not go out of their way to install the mechanisms on their devices if they were not installed by default. Among all mechanism categories, the icon overlay, which shows an alerting icon on the top of the screen to indicate bystanders, was most favoured by participants, followed by haptic feedback, physical tangible objects, and adjusting screen brightness to protect privacy. Further, we found that the general perception of mechanisms can be categorised into two components: perceived usefulness and personalisation of privacy protection mechanisms. Participants who viewed out-of-device privacy as highly important were more likely to favour the usefulness and personalisation of protection mechanisms. Participants with a high tendency to engage with technology scored low on the perceived usefulness of mechanisms but preferred personalisation more. Our findings also showed no differences in preferences for perceived usefulness and personalisation of mechanisms between iOS and Android users. 

Our research highlights how personal attributes shape preferences for protection against shoulder surfing. Our work develops a better understanding of user needs that guide towards user-tailored privacy. Based on the results, we present key takeaways to assist developers and researchers when designing protection mechanisms. \\

\noindent \textbf{Contribution Statement: }In this paper, we make the following contributions: 
\begin{enumerate}
    \item \textbf{Literature Review: }We present the results of a systematic literature review that narrows down the list of content-based protection mechanisms against shoulder surfing. 
    \item \textbf{User Perception Investigation: }We evaluate the users' perception of protection mechanisms extracted from the literature review to identify personal attributes' influence on preference and design of protection mechanisms. 
    \item \textbf{Key Takeaways: }Based on the results, we present key takeaways useful for designers and researchers for building novel mechanisms. 
\end{enumerate}

\section{Background \& Related Work}
This section presents an overview of related work and reflects on the background of content-based shoulder surfing. 
\subsection{User-Level \& Device-Level Protection Mechanisms}
Numerous studies focusing on shoulder surfing have provided evidence for users' concerns around shoulder surfing~\cite{farzand2021interplay,farzand2022shoulder,farzand2022hate,eiband2017understanding}. Studies report that users are concerned about their and other people's privacy who's data is viewed~\cite{eiband2017understanding}. Users also reported that shoulder surfing is not just a breach of privacy but also gives rise to negative feelings between the user and the observer~\cite{farzand2021interplay,eiband2017understanding,farzand2022shoulder}. Due to privacy and social concerns arising from shoulder surfing, users opt for various manual protection measures such as tilting the device, switching between apps, or turning off the phone~\cite{eiband2017understanding,farzand2022shoulder,farzand2021interplay}. Users have also reported using a privacy screen as a protective measure to safeguard their devices from shoulder surfing~\cite{eiband2017understanding}. Use of these measures has been shown as dependent on the relationship between the observer and the user~\cite{farzand2021interplay}. While these measures provide some basic protection against shoulder surfing, they are limited in their efficacy~\cite{khan2018evaluating}. Moreover, the user also needs to always remain alert to spot potential shoulder surfers and use user-level protection measures, which may not always be feasible. To overcome the limitations of user-level protection measures and provide an enhanced privacy experience, researchers have proposed several device-level mechanisms. Such mechanisms include lowering the screen brightness~\cite{zhou2016enhancing} or replacing the content displayed on the screen with randomly generated content~\cite{mitchell2015cashtags}. Researchers have also explored haptic-based mechanisms such as device vibrations to alert the user for shoulder surfing~\cite{saad2018communicating}. Related work pins down several similar mechanisms that either offer content-specific protection, such as for photos or text~\cite{vishwamitra2017blur} or provide full-screen protection regardless of content~\cite{khamis2018eyespot}. Despite the huge effort of researchers in developing novel ways to mitigate shoulder surfing and protect the privacy of users, the protection mechanisms proposed in the literature remain unexplored. This is important to know as it provides evidence of what protection ideas have been developed and evaluated and also paves the way for further improvement and refinement of those protection mechanisms. Having this knowledge allows researchers to expand on the existing ideas and research on undiscovered aspects of mitigating shoulder surfing.

\subsection{Personal Attributes \& Privacy Preferences}
Users vary in their needs and preferences for privacy and, therefore, need to be clustered based on their privacy profiles. One of the commonly used user profiling methods is Westin's three categories: unconcerned, fundamentalists, and pragmatists~\cite{kumaraguru2005privacy}. However, recent work has argued that Westin's three categories might not be related to users' corresponding behaviour~\cite{woodruff2014would,consolvo2005location}. In response to this criticism, Dupree et al. suggested five privacy personas (fundamentalists, lazy experts, technicians, amateurs, and marginally concerned)~\cite{dupree2016privacy}. While the method of Dupree et al. provided an overall picture of user profiles, researchers have also looked at profiling users for specific topics. For example, for smartphone privacy settings~\cite{frik2022users}, app permissions~\cite{lin2014modeling,liu2014reconciling}, location sharing~\cite{consolvo2005location}, or social media privacy behavior~\cite{wisniewski2017making}. In addition to profiling methods, researchers have also proposed various scales to capture granular concepts such as preferences or concerns to improve users' experience with technology. For example, Internet Users Information Privacy Concerns was proposed by Malhotra et al. to capture the privacy concerns of internet users~\cite{malhotra2004internet}. Hasan et al. proposed a psychometric scale to capture the importance of other people's privacy~\cite{hasan2023psychometric}.  For threats that exist outside the device, such as shoulder surfing, Farzand et al. proposed an out-of-device privacy scale to measure the importance users attribute towards protecting their information from out-of-device threats~\cite{farzand2024out}. While literature presents an enormous collection of profiling users based on their privacy preferences, expectations, and concerns, it remains unclear how these measurements correlate with users' preferences for privacy mechanisms such as age, gender, and the like.  

\subsection{Research Gap}
Extensive research has been carried out in exploring ways to mitigate shoulder surfing in the daily lives of users~\cite{zhou2016enhancing,mitchell2015cashtags,khamis2018eyespot}. However, the one-size-fits-all approach cannot be applied due to individual differences. To this end, we explore how personal attributes shape a user's preference for protection mechanisms against shoulder surfing.

\section{Stage 1: Collecting Content-Based Protection Mechanisms against Shoulder Surfing}\label{sec:sec1}
This section describes conducting the systematic literature review to answer \textbf{RQ$_1$}. Our search methodology was as follows: \\

\noindent\textbf{\textit{1) Keywords and Search Space:}} Two researchers from the field of Usable Security \& Privacy discussed the keywords to perform the search and finalized the search query: [All: "shoulder surfing"] OR [All: "shoulder surfing attack*"] OR [All: "shoulder-surfing attack*"] AND [E-Publication Date: (01/01/1999 TO *)]. For the search space, we selected the top 10 venues in "Human-Computer Interaction" and the top 10 in "Computers Security \& Cryptography" according to Google Scholar's ranking system (date assessed: July 2024). We selected the time frame of research publication after 1999 as 1999 is the year when one of the most influential human-centred security papers was published~\cite{adams1999users}. \\

\noindent\textbf{\textit{2) Exclusion Criteria:}} From the search results, we excluded a paper if it did not include the keywords in the full text. We also excluded a paper if it was a poster, a survey or a literature review paper. Further, we excluded a paper that presented an authentication-based mechanism as this paper focuses on mechanisms for content (such as text or photos) or targets a device other than a smartphone or a tablet. We included tablets as smartphones and tablets follow similar design principles. The main search resulted in five papers. \\

\noindent\textbf{\textit{3) Forward and Backward Search:}} For each paper identified in Step 2, we performed a backwards and forward search to identify relevant papers published elsewhere. This was done to ensure a broad coverage of research contributions in content-based shoulder surfing protection mechanisms. We then reapplied the exclusion criteria to the papers from the forward and backward search. Nine papers were identified through forward and backward search. Table~\ref{tab:venueresults} presents an overview of the search results. \\

\noindent\textbf{\textit{4) Extraction of Mechanisms:}} For each of the papers identified from the main search and extended search, we extracted the protection mechanisms. A total of 27 mechanisms were extracted. 

\begin{table*}[]
\centering
\resizebox{\textwidth}{!}{%
\begin{tabular}{@{}clcc@{}}
\toprule
\multicolumn{4}{c}{\textbf{Computers \& Security Top 10 Venues}}                                                                  \\ \midrule
\multicolumn{1}{l}{} & \multicolumn{1}{c}{Venue}                                    & Resulted Papers      & Selected Papers      \\
1                    & IEEE Symposium on Security and Privacy                       & 30                   & 0                    \\
2                    & USENIX Security Symposium                                    & 35                   & 2                    \\
3                    & IEEE Transactions on Information Forensics and Security      & 30                   & 0                    \\
4                    & Computers \& Security                                        & 48                   & 0                    \\
5                    & ACM Symposium on Computer and Communications Security        & 32                   & 0                    \\
6                    & Network and Distributed System Security Symposium (NDSS)     & 4                    & 0                    \\
7                    & IEEE Transactions on Dependable and Secure Computing         & 18                   & 0                    \\
8                    & Journal of Information Security and Applications             & 22                   & 0                    \\
9                    & International Conference on Theory and Applications of Cryptographic Techniques (EUROCRYPT) & 0            & 0           \\
10                   & international Cryptology Conference (CRYPTO)                 & 0                    & 0                    \\
\multicolumn{1}{l}{} &                                                              & \multicolumn{1}{l}{} & \multicolumn{1}{l}{} \\
\multicolumn{1}{l}{} & \multicolumn{1}{c}{\textbf{Total}}                           & \textbf{219}         & \textbf{2}           \\ \midrule
\multicolumn{4}{c}{\textbf{HCI Top 10 Venues}}                                                                                    \\ \midrule
\multicolumn{1}{l}{} & \multicolumn{1}{c}{Venue}                                    & Resulted Papers      & Selected Papers      \\
1                    & Computer Human Interaction (CHI)                             & 102                  & 2                    \\
2                    & Proceedings of the ACM on Human-Computer Interaction         & 9                    & 0                    \\
3                    & International Journal of Human-Computer Studies             & 10                   & 0                    \\
4                    & International Journal of Human-Computer Interaction          & 12                   & 0                    \\
5                    & IEEE Transactions on Affective Computing                     & 0                    & 0                    \\
6                    & Behaviour \& Information Technology                          & 7                    & 0                    \\
7                    & Virtual Reality                                              & 18                   & 0                    \\
8                    & Proceedings of the ACM on Interactive, Mobile, Wearable and Ubiquitous Technologies         & 20           & 0           \\
9                    & International Journal of Interactive Mobile Technologies     & 26                   & 1                    \\
10                   & ACM/IEEE International Conference on Human Robot Interaction & 0                    & 0                    \\
\multicolumn{1}{l}{} &                                                              & \multicolumn{1}{l}{} & \multicolumn{1}{l}{} \\
\multicolumn{1}{l}{} & \multicolumn{1}{c}{\textbf{Total}}                           & \textbf{204}         & \textbf{3}           \\
\multicolumn{1}{l}{} &                                                              & \multicolumn{1}{l}{} & \multicolumn{1}{l}{} \\ \midrule
\multicolumn{1}{l}{} & \multicolumn{1}{c}{\textit{\textbf{Forward \& Backward Search (N=5)}}}                      & \textbf{262} & \textbf{9} \\ \bottomrule
\end{tabular}%
}
\caption{The Table shows the results of the systematic literature review conducted on the top 10 venues in HCI and Computer Security according to Google Scholar (accessed: July 2024). }
\label{tab:venueresults}
\end{table*}

\section{Stage 2: Categorisation of Mechanisms}
After identifying the mechanisms against shoulder surfing in the literature, the next step was categorising them into groups. This section explains our rationale and categorisation strategy for the mechanisms. 

\subsection{Categorisation Strategy}
In this stage, we categorised the mechanisms resulting from Stage~\ref{sec:sec1} based on how the information is displayed on the screen. For this, one researcher extracted all mechanisms proposed in each of the papers along with their pictorial representations and descriptions. Then, the researcher identified commonalities in the design of information presentation in the presence of a shoulder surfer. Mechanisms with similar designs were grouped under one category. This resulted in a total of ten mechanism categories. Table~\ref{tab:mechanisms_categories} presents each mechanism's categorisation and source. A group description for each of the mechanism categories was then formulated. Next, two researchers reviewed the categorisation and the group descriptions, and any disagreements were resolved in a meeting. The categorisation and descriptions were revised based on the discussion. Next, we created videos of all mechanisms showcasing their functionality and included them as a group mechanism in the survey. To ensure an accurate reflection of each mechanism, we first checked for if a video representation of the mechanism had been made available by the researchers. If it was available, we used the respective mechanism video. In case a video was not available, we relied on the information available in the paper to develop a video prototype. This ensured that participants received a complete and accurate idea of all mechanisms in each category.

\begin{table*}[]
\centering
\resizebox{\textwidth}{!}{%
\begin{tabular}{@{}cp{0.7\textwidth}c@{}}
\toprule
\textbf{Mechanism Category} &
  \multicolumn{1}{c}{\textbf{Description}} &
  \textbf{Source} \\ \midrule
Display Color Change &
  a mechanism that changes the display into a monochrome combination of black and white upon the detection of bystanders &
  \cite{zhou2016enhancing} \\
Screen Brightness &
  a mechanism that lowers the screen brightness to make the content less visible from a distance &
  \cite{zhou2016enhancing,zhang2023don,lian2013smart,chen2019keep} \\
Selective Visibility &
  a mechanism that offers a selective viewing experience by making the part the user gazes at visible while making the rest invisible to the human eye &
  \cite{zhou2016enhancing,von2016you,khamis2018eyespot} \\
Distortion &
  a mechanism that obfuscates the image into small pixels &
  \cite{zhou2016enhancing,tang2023eye,khamis2018eyespot,vishwamitra2017blur,tajik2019balancing} \\
Blurry &
  a mechanism that makes the device screen unclear, reducing the sharpness of the display &
  \cite{tang2023eye,vishwamitra2017blur,darling2022privacy} \\
Replacement by Protective Text &
  A mechanism that replaces the content of the device screen by randomly generated unmeaningful text is activated. &
  \cite{mitchell2015cashtags} \\
Replacement by Meaningful Text &
  a mechanism replaces the content of the device screen with randomly generated meaningful text &
  \cite{khamis2018eyespot} \\
Physical Tangible Component &
  a mechanism that relies on physical components, i.e. external to the device screen itself, not involving rendering on display - is activated, such as an LED light, which is an external part of the device but not the screen display. &
  \cite{saad2018communicating} \\
Icon Overlay &
  an icon-like mechanism that is placed on the top of the screen to indicate bystanders &
  \cite{saad2018communicating} \\
Haptic &
  A mechanism that relates to the sense of touch is activated. In the case of shoulder surfing, the user will be alerted through phone vibrations &
  \cite{saad2018communicating} \\
Combination &
  A mechanism that combines multiple features into one mechanism is activated upon the detection of a shoulder surfer. For example, it adjusts font size, color scheme, and screen brightness - all of them together. &
  \cite{lei2021iscreen,ragozin2019private} \\ \bottomrule
\end{tabular}%
}
\caption{The Table shows an overview of the ten mechanism categories derived from literature. }
\label{tab:mechanisms_categories}
\end{table*}

\section{Stage 3: Data Collection}
After identifying and categorising the protection mechanisms in the literature, our next step was to develop a questionnaire and collect data. This section details our methodology and data analysis and presents the questionnaire's results.

\subsection{Questionnaire Design}
Our goal was to collect user preferences for protection mechanisms against content-based shoulder surfing. To collect data from a large and diverse population, we developed an online questionnaire on Qualtrics~\cite{Qualtrics}. The questionnaire comprised of the following components: 

\begin{enumerate}
    \item \textbf{Information Sheet \& Consent: }Before beginning the questionnaire, we presented participants with an information sheet detailing the aim and required tasks for the completion of the study. We presented the participants with information on shoulder surfing with a detailed description and pictorial representation. The participants were then asked to electronically sign the consent form and confirm they were aged 18+ if they wished to proceed. 
    \item \textbf{Demographics: }We asked our participants to indicate their age, gender, education, employment status, daily usage of smartphones (in hours), and the smartphone OS they used. 
    \item \textbf{Importance for Privacy: }Privacy preferences were captured through collecting importance for out-of-device privacy~\cite{farzand2024out}. The out-of-device privacy scale captures the importance an individual attributes towards protecting their information from out-of-device threats. This contrasts online information privacy concerns captured through scales such as IUIPC~\cite{malhotra2004internet}, which specifically focus on online information. Since shoulder surfing is one of the out-of-device threats, we selected ODPS to capture users' importance of privacy. 
    \item \textbf{Familiarisation with Technology: }To assess users' tendency to actively engage in or avoid intensive technology interaction, we used Affinity for Technology Scale~\cite{franke2019personal}. 
    \item \textbf{Mechanism-Related Questions: }Each mechanism category was first introduced with a description and video representation of all mechanisms that fall under the mechanism category. Participants were asked to focus on the presentation of the information in the presence of a shoulder surfer and then answer the follow-up questions. In the follow-up questions, we asked participants questions around (1) use, (2) importance, (3) protection, (4) preference for the mechanism or similar digital alternative, and (5) preference for a non-digital alternative in comparison to the presented mechanism. Each question was presented on a 7-point Likert scale (1=strongly disagree to 7=strongly agree).
    \item \textbf{General Items: }In addition to mechanism-specific questions, we also asked participants questions about (1) perceived usefulness and (2) trust in privacy protection mechanisms. Each question was presented on a 7-point Likert scale (1=strongly disagree to 7=strongly agree).
\end{enumerate}

\subsection{Pilot Study}
The questionnaire was first internally tested by N=7 experts with expertise in different domains under human-computer interaction. Experts provided feedback mainly around the wording of questions. Following the feedback from experts, it was then tested externally with N=10 participants recruited through Prolific. Participants were asked to report any issues related to understanding the questions or playing the videos in an open-ended question. They were also asked to indicate their browser-in-use to understand if playing the videos was an issue specific to a browser. Participants did not experience any issues in playing the videos. However, a few participants mentioned some ambiguities in the checks we had placed to ensure video watching. The questions were improved based on participants' feedback. All participants were compensated as per Prolific's recommendation. Overall, pilot testing with experts and with the target audience ensured that our study was focused on the research questions and was precisely understood by the general audience. 

\subsection{Ethical Considerations}
Our study was approved by the Ethics Committee at our institute. The study started with presenting an information sheet to participants detailing the study's aim, data collection, data usage, and data protection. We then presented them with a consent form to electronically sign if they wished to proceed. Participants had to be above the age of 18 to be eligible to participate in the study. We also turned off recording all respondents' IP addresses and location data collected through Qualtrics. To preserve anonymity, we deleted the Prolific's IDs after compensating all participants. Participants were compensated £8.08 per hour for their time and participation.

\subsection{Data Quality Checks}
Online data collection brings the advantage of collecting data from a large and diverse sample. However, it also brings the challenge of ensuring good-quality data collection. For this purpose, we undertook various measures such as (1) multiple \textbf{attention checks} in the questionnaire to check if participants were paying attention. We used the following attention check: \\

\textit{Research shows that some participants in online studies do not pay attention. Please help us monitoring the quality of our study results by answering this question with somewhat disagree.}\\

(2) We used \textbf{comprehension checks} to ensure that the participant had correctly understood the terminology and topic under investigation. In case of incorrect answers, participants were allowed to reread the information provided and then answer the questions again. As part of comprehension checks, participants were asked to indicate if the following statements were true or false: \\

(a) \textit{Shoulder surfing refers to observing someone's device screen without permission. Please indicate if this statement is true or false.} (b) \textit{To perform a shoulder surfing attack, one must be close to the device screen being observed. Please indicate if this statement is true or false.} \\

(3) To ensure that participants viewed all videos in each mechanism category, we asked \textbf{an additional question} after the videos were showcased related to the content shown in the videos. If participants answered incorrectly, they were not allowed to proceed but were asked to rewatch the videos and re-answer the questions. An example of such a question is \\

\textit{The above videos show examples of screen brightness for ....} \\

Participants could then choose between (i) text messages only, (ii) text messages, chart visualisation, and photos, and (iii) text messages and photos based on the content shown in the videos. (4) Finally, we utilised the feature of \textbf{bot detection} provided by Qualtrics to avoid unwanted submissions. 

\section{Results}
The goal of this study is to understand user Personalisation mechanisms against shoulder surfing. In this section, we present the results of user preferences for ten protection mechanism categories evaluated with a UK-based population. 

\subsection{Recruitment \& Participants}
We recruited N=226 participants from the UK through Prolific. Twenty-five participants' data was removed as they failed to pass the attention checks. On average, participants took 25 minutes to complete the questionnaire (sd=11.20). Participants who took less than half the average time were removed from the analysis (N=9). No submission was marked as a bot submission. Out of the remaining N=192 participants, N=93 self-identified as man, N=95 as woman, N=3 as non-binary, and one participant preferred not to say. Participants' mean age was 42.78 (sd=13.94, min=18, max=78). More than half of the participants were employed full-time (N=103), N=34 were employed part-time, and N=20 were unemployed. Fourteen participants were retired, N=9 homemakers, and N=8 were students. Most participants held a 4-year degree (N=61) or a professional degree (N=42), followed by some college (N=40). Twenty-nine participants were high school graduates, N=10 held two-year degrees, and a few (N=5) had doctorates or had education less than high school. Most participants (N=96) reported to spend between 2 and 5 hours per day interacting with smartphones. A majority of participants (N=44) spent less than 2 hours, N=33 spent between 5 and 7 and a small group (N=19) spent more than 7 hours per day interacting with smartphones. A vast majority of participants were Android smartphone users (N=113), while N=79 participants were iOS users.

\subsection{Experience with Shoulder Surfing}
More than half of the participants agreed that they had been shoulder surfed (N=113), while a large group of participants could not recall (N=57), and a small group of participants disagreed (N=22). When eliciting the specific role in a shoulder surfing situation, almost half of the participants agreed that they experienced shoulder surfing as a victim and also as an observer (N=80). Some participants shared they had shoulder-surfed someone (N=44), they were shoulder-surfed by someone (N=39), and a few participants preferred not to answer (N=29). Thirty-six participants recalled having the latest experience of shoulder surfing in the last six months, N=35 participants never had an experience, N=30 experienced in the last few days, followed by more than a year ago (N=27) and in the last three months (N=26). Twenty-three participants experienced it last month, and N=15 experienced it a few weeks ago. Almost all participants (N=175) held the opinion that they are responsible for protection against shoulder surfing, and a few (N=13) expressed that both the manufacturer and the user are responsible for protecting against it. Only a few participants felt that the device manufacturer was responsible for protection against shoulder surfing (N=3). 

\subsection{Perceptions of Protection Mechanisms}

\textbf{Overall Perception: }Overall, our participants slightly agreed that the mechanisms would help in protecting their privacy (protection.median= 5, st dev=1.66, mean=4.34), but they also held the opinion that they would prefer having non-digital alternatives than the presented mechanisms (alternate.median= 5, st dev=1.64, mean=4.76). When enquired about if participants would use the mechanism, would like to have it installed on their phone, or if not installed, then they would install it or look for an alternate, participants slightly disagreed (availability.median = 3, st dev =1.78, mean=3.26). Participants held similar views regarding whether they would like to use the protection mechanisms (use.median=3, st dev=1.82, mean=3.5) and if it's important for them to have the mechanism installed on their devices (own.median=3, st dev=1.74, mean=3.35).

When comparing individual mechanisms, our participants somewhat agreed that they would prefer using non-digital alternatives in comparison to all the presented mechanisms, such as covering the screen using their hands (alternate.{median= 5}). Participants disagreed or somewhat agreed with installing the mechanism on their phones if not installed by default for all mechanisms except haptic and icon overlay, where participants were found to be neutral (availability.median = 4). Similar views were seen when asked if participants would prefer installing the mechanism and working on their devices. Most mechanisms were not favoured, and participants disagreed or somewhat disagreed except for icon overlay and haptic, where participants were found to be neutral (own.median = 4). Participants expressed agreement for using icon overlay to protect their privacy (use.median=5), whereas they were found to be neutral for tangible and haptic (own.median=4) and disagreed for all the rest of the mechanisms (use.median=3-2). 


\begin{figure*}
    \centering
\includegraphics[width=\textwidth]{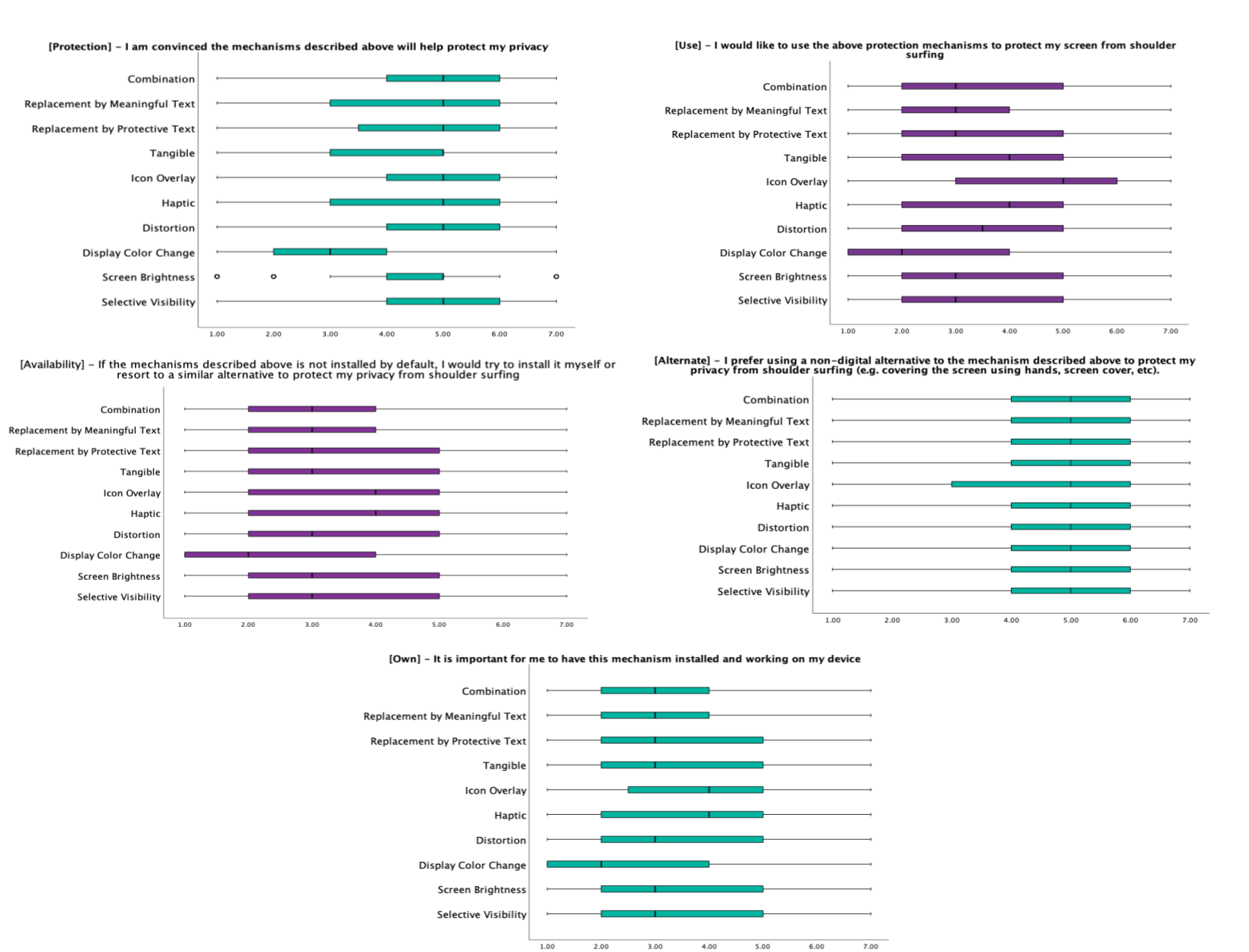}
    \caption{The Figure shows participants’ feedback on each protection mechanism category. Participants could select from a 7-point Likert scale ranging from strongly disagree (1) to strongly agree (7). }
    \label{fig:enter-label}
\end{figure*}

\textbf{Similarities \& Differences between Mechanisms: }
We next analysed the distance between mechanisms to see if mechanisms were perceived as similar or different. For this, we calculated Euclidean distance (d) between all mechanisms' mean responses. The smallest distance (d=11.221) was seen between replacement by meaningful text and combination mechanisms. Similarly, smaller distances were observed between distortion, selective visibility, replacement by meaningful text, replacement by protective text, and combination (12.207 $<$ d $<$ 12.787). On the contrary, the largest distance was observed between display colour change and icon overlay (d=22.631). Distances close to each other indicate that mechanisms were perceived similarly, whereas larger distances indicate that the perception of mechanisms differed greatly. Table~\ref{tab:distances} presents the detailed results.

\subsection{Ranking of Mechanisms}
\begin{figure*}
    \centering
    \includegraphics[width=\textwidth]{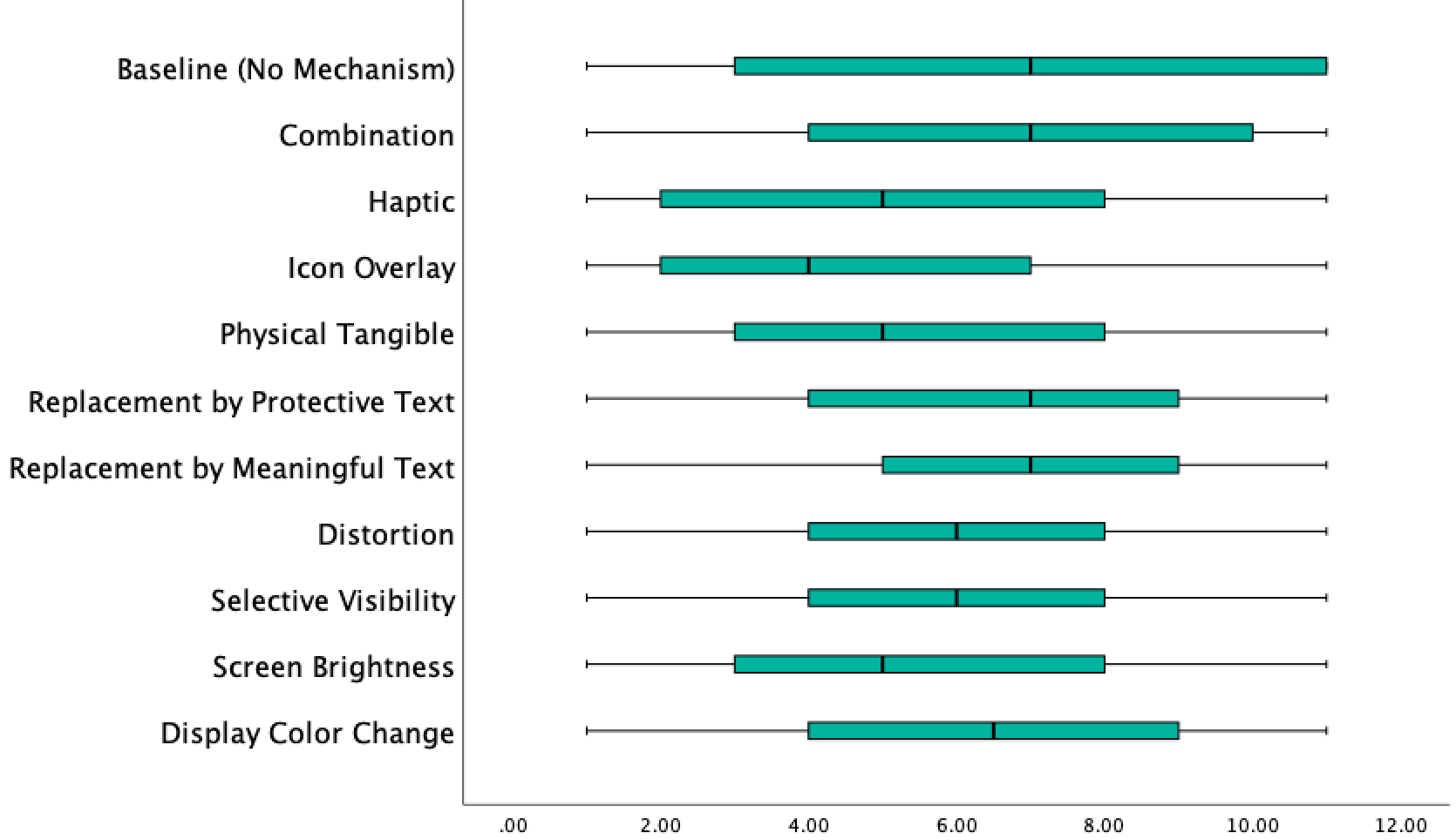}
    \caption{The figure shows the ranking of protection mechanism categories.}
    \label{fig:ranking}
\end{figure*}

Figure~\ref{fig:ranking} shows participants' ranking of the ten mechanism categories. Icon overlay was most favoured (median=4), followed by haptic, physical tangible, and screen brightness (median=5). All remaining mechanisms, including baseline (i.e. no protection mechanisms), were least preferred by participants ( 6 $<$ median $<$ 7 ). A Friedman test found significant differences between the ranking of mechanisms (x2 (10) =102.063, \textit{p} $<$ 0.001). Pairwise comparisons were performed with a Bonferroni correction for multiple comparisons. Rankings were statistically significant between different mechanisms. Post hoc analysis revealed significant differences between (1) icon overlay and distortion, selective visibility, combination, display color change, replacement by protective text, replacement by meaningful text, and baseline (no protection), (2) physical tangible and replacement by meaningful text and baseline (no protection), (3) haptic and combination, display colour change, replacement by protective text, replacement by meaningful text and baseline (no protection), and (4) icon overlay and distortion, selective visibility, combination, display color change, replacement by protective text, replacement by meaningful text and baseline (no protection). 

\subsection{General Perception of Mechanisms}
After assessing the presented mechanisms, we looked into the overall perception of protection mechanisms by asking questions about (1) Personalisation and (2) Perceived Usefulness. 

Participants agreed that an understanding of these mechanisms to protect their privacy was easy (median=6). Participants also somewhat agreed (median =5) that protecting text and photos from shoulder surfing was important to them. They also felt that they would be more comfortable using their smartphones in public if the mechanisms were installed on their phones, which would help safeguard their privacy (median =5). Participants also expressed trust in mechanisms for protecting their privacy, and having access to mechanisms would make them consider more about their privacy and be aware of their surroundings (median =5). They would prefer having mechanisms on their phone rather than none, and the presented mechanisms better protect their privacy than purely non-digital alternatives such as covering my hands (median =5). Participants also felt that having a mechanism that only alerts them about shoulder surfing is sufficient for them (median =5). Participants were found to be neutral that the mechanism must convey irrelevant information to the observer to let them know they have invaded their personal space (median =4). Furthermore, they also held a neutral opinion that the protection mechanism should tell about who the observer is, but simultaneously, the mechanism should not let the observer know that the user has a mechanism activated on their phones (median =4). Lastly, participants were also neutral on having a mechanism that covers the entire display (median =4). Table~\ref{tab:overall_perceptions_of_protection_mechanisms} shows the descriptive results for participants' perspectives on general mechanisms.

\begin{table*}[]
\centering
\resizebox{\textwidth}{!}{%
\begin{tabular}{@{}p{.7\textwidth}ccc@{}}
\toprule
\textbf{Statements} &
  \textbf{Median} &
  \textbf{Mean} &
  \textbf{Standard Deviation} \\ \midrule
Protecting text and photos from shoulder surfing is important to me. &
  5 &
  4.62 &
  1.63 \\
I would feel more comfortable using my smartphone in public if it has privacy protection mechanisms. &
  5 &
  4.61 &
  1.62 \\
Having protection mechanisms will safeguard my privacy. &
  5 &
  5.20 &
  1.32 \\
Understanding how these mechanisms protect my privacy is easy. &
  6 &
  5.53 &
  1.28 \\
I trust the presented mechanisms to protect my privacy from shoulder surfing. &
  5 &
  4.55 &
  1.43 \\
Having these mechanisms makes me aware of my surroundings &
  5 &
  5.11 &
  1.26 \\
Having access to the mechanisms described above makes me consider using them to protect my privacy. &
  5 &
  4.62 &
  1.68 \\
I prefer using the presented mechanisms rather than having none to protect my device from shoulder surfing. &
  5 &
  4.37 &
  1.73 \\
The presented privacy mechanisms better protect my privacy from shoulder surfing than non-digital alternatives, such as covering the screen using my hands or screen cover. &
  5 &
  4.51 &
  1.57 \\
Having a mechanism that only alerts me about shoulder surfing is sufficient for me. &
  5 &
  4.89 &
  1.38 \\
The protection mechanism must convey irrelevant information (such as random unmeaningful text)  to the observer to let them know they have invaded my personal space. &
  4 &
  3.91 &
  1.62 \\
A protection mechanism that covers the entire display is suitable for me. &
  4 &
  4.09 &
  1.61 \\
The protection mechanism should tell about who the observer is as well. &
  4 &
  3.80 &
  1.50 \\
The observer should not know that I have a protection mechanism. &
  4 &
  4.29 &
  1.68 \\ \bottomrule
\end{tabular}%
}
\caption{The Table shows the overall perception of protection mechanisms (general)}
\label{tab:overall_perceptions_of_protection_mechanisms}
\end{table*}

\begin{table*}[]
\centering
\resizebox{\textwidth}{!}{%
\begin{tabular}{@{}p{0.2\textwidth}ccccccccc@{}}
\toprule
{\color[HTML]{264A60} } &
  { \textbf{Selective Visibility}} &
  { \textbf{Screen Brightness}} &
  {\textbf{Display Color Change}} &
  {\textbf{Distortion}} &
  { \textbf{Haptic}} &
  { \textbf{Icon Overlay}} &
  { \textbf{Tangible}} &
  {\textbf{Rep. by Protective Text}} &
  { \textbf{Rep. by Meaningful Text}} \\ \midrule
{\textbf{Screen Brightness}} &
  \cellcolor[HTML]{D4EBDC}14.78 &
   &
   &
   &
   &
   &
   &
   &
   \\
{\textbf{Display Color Change}} &
  \cellcolor[HTML]{F9FAFC}15.949 &
  \cellcolor[HTML]{FCEEF1}16.664 &
   &
   &
   &
   &
   &
   &
   \\
{\textbf{Distortion}} &
  \cellcolor[HTML]{94D2A5}12.787 &
  \cellcolor[HTML]{D2EBDA}14.72 &
  \cellcolor[HTML]{FCECEF}16.773 &
   &
   &
   &
   &
   &
   \\
{\textbf{Haptic}} &
  \cellcolor[HTML]{FCF2F5}16.506 &
  \cellcolor[HTML]{FBB5B7}19.256 &
  \cellcolor[HTML]{FA9EA1}20.254 &
  \cellcolor[HTML]{F2F8F6}15.729 &
   &
   &
   &
   &
   \\
{\textbf{Icon Overlay}} &
  \cellcolor[HTML]{FCE5E8}17.084 &
  \cellcolor[HTML]{FBC0C3}18.753 &
  \cellcolor[HTML]{F8696B}22.631 &
  \cellcolor[HTML]{FCF3F6}16.465 &
  \cellcolor[HTML]{C5E5CF}14.319 &
   &
   &
   &
   \\
{ \textbf{Tangible}} &
  \cellcolor[HTML]{FCF7F9}16.302 &
  \cellcolor[HTML]{FBC2C5}18.654 &
  \cellcolor[HTML]{FAB1B4}19.406 &
  \cellcolor[HTML]{FCE4E7}17.128 &
  \cellcolor[HTML]{E1F1E7}15.191 &
  \cellcolor[HTML]{CDE9D6}14.564 &
   &
   &
   \\
{ \textbf{Rep. by Protective Text}} &
  \cellcolor[HTML]{DCEFE3}15.039 &
  \cellcolor[HTML]{FCD8DB}17.678 &
  \cellcolor[HTML]{FAAFB2}19.507 &
  \cellcolor[HTML]{C0E3CB}14.165 &
  \cellcolor[HTML]{FCFCFF}16.034 &
  \cellcolor[HTML]{FCF9FC}16.2 &
  \cellcolor[HTML]{FCDEE1}17.401 &
   &
   \\
{\textbf{Rep. by Meaningful Text}} &
  \cellcolor[HTML]{C2E4CD}14.22 &
  \cellcolor[HTML]{FCFCFF}16.072 &
  \cellcolor[HTML]{F0F7F5}15.687 &
  \cellcolor[HTML]{8DCF9F}12.549 &
  \cellcolor[HTML]{FCFBFE}16.08 &
  \cellcolor[HTML]{FCEFF2}16.62 &
  \cellcolor[HTML]{F3F8F8}15.78 &
  \cellcolor[HTML]{85CC98}12.311 &
   \\
{\textbf{Combination}} &
  \cellcolor[HTML]{9CD5AC}13.022 &
  \cellcolor[HTML]{DAEEE1}14.969 &
  \cellcolor[HTML]{FCEDF0}16.731 &
  \cellcolor[HTML]{82CA96}12.207 &
  \cellcolor[HTML]{EEF6F3}15.623 &
  \cellcolor[HTML]{FCF9FC}16.204 &
  \cellcolor[HTML]{EAF4EF}15.473 &
  \cellcolor[HTML]{ABDBB9}13.492 &
  \cellcolor[HTML]{63BE7B}11.221 \\ \bottomrule
\end{tabular}%
}
\caption{Distances between Mechanisms indicating similarities and differences}
\label{tab:distances}
\end{table*}

We then conducted a principal component analysis to determine if multiple privacy items are related to the same factors. A principle component analysis with oblique promax rotation resulted in three factors, one of which had only two items loaded onto it. A visual inspection of the scree plot also suggested factors between two or three. Before finalising the solution, we calculated reliability using Cronbach's alpha, which resulted in poor reliability for the third factor (which had only two items). Further, considering the best principles and recommendations around dimension reduction, we explored two-factor solutions. The two-factor solution resulted in an improved, simple solution. We then again checked for reliability, and in the case of a two-factor solution, appreciable reliability was achieved for both factors ($>$ 0.70). We named the factors "Perceived Usefulness" ($\alpha$ = 0.797) and "Personalisation" ($\alpha$ = 0.876). Perceived usefulness represents if users do prefer having the mechanism to protect from shoulder surfing. Next to determining if the participants want a mechanism, the component "Personalisation" captures how the user wants the mechanism to be that is tailored to their preferred personalisation. 

Table~\ref{tab:PCA} presents the results of the two-factor solution of principle component analysis.

\begin{table*}[]
\centering
\resizebox{\textwidth}{!}{%
\begin{tabular}{@{}cp{0.6\textwidth}ll@{}}
\toprule
\multicolumn{1}{l}{} &
   &
  \multicolumn{2}{c}{\textbf{Loadings}} \\ \midrule
\rowcolor[HTML]{CDF4BA} 
\cellcolor[HTML]{CDF4BA} &
  Having a mechanism that only alerts me about shoulder surfing is sufficient for me &
  -0.708 &
  0.847 \\
\rowcolor[HTML]{CDF4BA} 
\cellcolor[HTML]{CDF4BA} &
  Having protection mechanisms will safeguard my privacy &
   &
  0.809 \\
\rowcolor[HTML]{CDF4BA} 
\cellcolor[HTML]{CDF4BA} &
  Understanding how these mechanisms protect my privacy is easy &
   &
  0.78 \\
\rowcolor[HTML]{CDF4BA} 
\cellcolor[HTML]{CDF4BA} &
  I trust the presented mechanisms to protect my privacy from shoulder surfing &
   &
  0.717 \\
\rowcolor[HTML]{CDF4BA} 
\cellcolor[HTML]{CDF4BA} &
  Having these mechanisms makes me aware of my surroundings &
   &
  0.609 \\
\rowcolor[HTML]{CDF4BA} 
\multirow{-6}{*}{\cellcolor[HTML]{CDF4BA}\textbf{Percieved Usefulness}} &
  The presented privacy mechanisms better protect my privacy from shoulder surfing than non-digital alternatives, such as covering the screen using my hands or screen cover &
  0.415 &
  0.426 \\
\rowcolor[HTML]{CBCEFB} 
\cellcolor[HTML]{CBCEFB} &
  The protection mechanism must convey irrelevant information (such as random unmeaningful text)  to the observer to let them know they have invaded my personal space. &
  0.763 &
   \\
\rowcolor[HTML]{CBCEFB} 
\cellcolor[HTML]{CBCEFB} &
  Protecting text and photos from shoulder surfing is important to me. &
  0.711 &
   \\
\rowcolor[HTML]{CBCEFB} 
\cellcolor[HTML]{CBCEFB} &
  A protection mechanism that covers the entire display is suitable for me. &
  0.709 &
   \\
\rowcolor[HTML]{CBCEFB} 
\cellcolor[HTML]{CBCEFB} &
  I prefer using the presented mechanisms rather than having none to protect my device from shoulder surfing. &
  0.681 &
   \\
\rowcolor[HTML]{CBCEFB} 
\cellcolor[HTML]{CBCEFB} &
  I would feel more comfortable using my smartphone in public if it has privacy protection mechanisms. &
  0.678 &
   \\
\rowcolor[HTML]{CBCEFB} 
\cellcolor[HTML]{CBCEFB} &
  The protection mechanism should tell about who the observer is as well. &
  0.617 &
   \\
\rowcolor[HTML]{CBCEFB} 
\multirow{-7}{*}{\cellcolor[HTML]{CBCEFB}\textbf{Personalisation}} &
  Having access to the mechanisms described above makes me consider using them to protect my privacy. &
  0.549 &
  0.431 \\
\multicolumn{1}{l}{\cellcolor[HTML]{FFFFFF}} &
  {\color[HTML]{FE0000} The observer should not know that I have a protection mechanism.} &
   &
   \\ \bottomrule
\end{tabular}%
}
\caption{The Table shows the result of the Principle Component Analysis representing components structure. The statement highlighted in red was removed as it did not load sufficiently high ($>$ 0.30).}
\label{tab:PCA}
\end{table*}

\subsection{Correlation between Personal Attributes \& Perception of Protection Mechanisms}
\textbf{Out-of-Device Privacy Scale: }First, we analysed the reliability of Out-of-Device Privacy Scale which resulted in optimum reliability ($\alpha$ = 0.918). Next, we analysed the link between ODPS and components of protection mechanisms. There was a strong positive association between ODPS and Personalisation, which was statistically significant ($\tau$ = 0.549, \textit{p} $<$ 0.001). ODPS had a weak positive association with Perceived Usefulness, which was statistically significant ($\tau$ = 0.300, = \textit{p} $<$ 0.001). 


\textbf{Affinity for Technology Interaction: }Acceptable internal consistency was observed across the dataset of Affinity for Technology Interaction scale ($\alpha$ = 0.891). There was a strong positive correlation between Personalisation and ATI ($\tau$ =0.105, \textit{p} = 0.037), but a weak association was observed with Perceived Usefulness which was not found to be statistically significant ($\tau$ = .109, \textit{p} = 0.032). This shows that users with high ATI scores prefer mechanisms that raise Personalisation, but the ATI score does not impact Perceived Usefulness.


\textbf{Age: }All participants shared a similar out-of-device privacy scale (ODPS) score regardless of the age group. Participants between the ages of 18 and 30 had a median score of 5.11 for Out-of-Device Privacy; participants between the ages of 31 and 60 also had a median score of 5.11, and participants between the ages of 61 and 78 had a median score of 5.17. Next, we calculated Kendall's tau correlation to test the relationship between ODPS and age. The results showed a weak correlation ($\tau$ =0.028) but were not found to be significant (\textit{p}=0.566). We next compared age with Personalisation ($\tau$=0.030, \textit{p}=0.545) and Perceived Usefulness ($\tau$=-0.037, \textit{p}=0.457) but did not observe significant differences. This shows that age does not play a part in forming preferences for protection mechanisms and neither in the level of concern for out-of-device privacy.

\textbf{Gender: }
Table~\ref{tab:gender} presents a descriptive summary of participants' perceptions of general privacy mechanisms. The mean responses for both components of general privacy mechanisms were found to be close between male and female subsets. We compared the subset of male and female responses for the two components of general privacy mechanisms. We found Kendal Tau's correlation for Personalisation ($\tau$ = 0.020, \textit{p} = 0.782) and for Perceived Usefulness ($\tau$ = 0.036, \textit{p} = 0.630). None of the correlations was found to be significant. This shows that there are no differences in preferences among users based on gender. 


\begin{table}[]
\centering
\begin{tabular}{@{}p{0.15\columnwidth}ccccccc@{}}
\toprule
 & \multicolumn{1}{l}{} & \multicolumn{3}{c}{\textbf{Personalisation}} & \multicolumn{3}{c}{\textbf{Perceived Usefulness}} \\ \midrule
gender            & n  & mean & st dev & median & mean & st dev & median \\
man               & 93 & 4.16 & 1.27   & 4.14   & 4.94 & 1.01   & 5.17   \\
woman             & 95 & 4.43 & 1.19   & 4.43   & 4.99 & 0.94   & 5      \\
prefer not to say & 1  & 4.71 & -      & -      & 5.5  & -      & -      \\
non-binary        & 3  & 3.38 & 1.39   & 3.71   & 4.78 & 1.17   & 4.67   \\ \bottomrule
\end{tabular}
\caption{The Table shows Participants' perception of general privacy mechanisms in relation to their gender. }
\label{tab:gender}
\end{table}


\textbf{Device Operating System: }
Among our sample, N=79 participants were iOS users, while most were Android users (N=113). We then assessed the relationship between the use of smartphone OS and components of general privacy protection mechanisms. We observed a negative association between iOS and Android users for Personalisation, but it was not found to be significant ($\tau$ = -0.030, \textit{p} =.702). On the contrary, no correlation was found between Android and iOS users for Perceived Usefulness ($\tau$ = 0, \textit{p}=.997). 


\section{Discussion}
Our goal was to assess the role of personal attributes in forming preferences for privacy protection mechanisms against shoulder surfing. In this section, we discuss key takeaways based on the results and propose future work directions. 

\subsection{Privacy Mechanisms for General Population}
Overall, our participants agreed that the presented mechanisms would offer protection, but they also expressed that they would prefer non-digital alternatives as opposed to device-based mechanisms. Participants also shared that they would not go out of their way to have the mechanism installed on their devices except for icon overlay or haptics. This shows that mechanisms should be available by default, and the user should not have to search or install mechanisms. Holistically, users perceived the understanding of how the protection mechanisms functioned as easy. Despite the easy understanding of the mechanisms' functionality, participants generally favoured icon overlay, haptic, and tangible mechanisms among all mechanisms. Echoing previous work, this finding shows that participants are more inclined towards alerting and unobtrusive mechanisms than mitigation mechanisms~\cite{farzand2021interplay,farzand2022shoulder}. This indicates that future design of protection mechanisms should consider mechanisms similar to icon overlay, haptics, or tangible mechanisms.

\subsection{Key Takeaways} \label{design_guidelines}

\begin{itemize}
    \item \includegraphics[height=20pt,width=20pt]{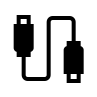} \textbf{Design Equity regardless of Smartphone OS: }Android and iOS are the two popular operating systems with different ecosystems. Research on privacy concerns of iOS and Android users reveals that none of the architectures is a winner regarding privacy~\cite{kollnig2022iphones}. Similar to other privacy concerns, we did not observe differences between iOS and Android users for perceived usefulness and personalisation of protection mechanisms against shoulder surfing. Based on this, designers can develop mechanisms without considering the smartphone OS specifications or target audience. \\
    \item \includegraphics[height=20pt,width=20pt]{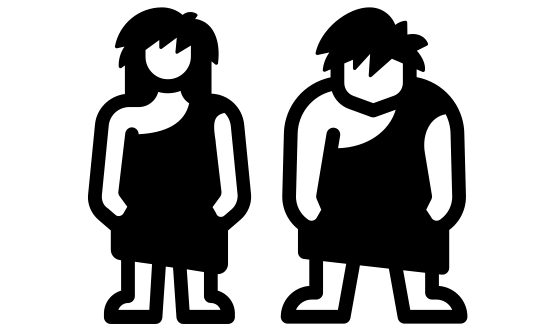} \textbf{Age-Independent Mechanisms: }Level of privacy concerns change as we move across different age groups, for example, young adults are often seen as engaging more in privacy-protective behaviours while older adults are seen as concerned for other individuals~\cite{kezer2016age}. However, in our study, we observed that age does not play a part in forming preferences for perceived usefulness and personalisation of mechanisms. This shows that while designing mechanisms, age does not play the role of a confounding variable. \\
    \item \includegraphics[height=20pt,width=20pt]{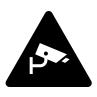} \textbf{User-Tailored Privacy: }Our results showed a positive and significant correlation between the Out-of-Device Privacy score and perceived usefulness and personalisation. This shows that researchers should incorporate user-tailored privacy, which recommends analysing user privacy profiles and then recommending privacy solutions. This finding echoes prior work that advocates for user-tailored privacy~\cite{knijnenburg2017privacy,knijnenburg2022user}. \\
    \item \includegraphics[height=20pt,width=20pt]{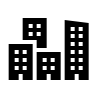} \textbf{Increased Personalisation with Increased Affinity for Technology Interaction: }In our study, users that had a high tendency to engage with technology-preferred mechanisms preferred personalization of mechanisms~\cite{franke2019personal}. This shows that tech-savvy users prefer mechanisms that they can adjust and modify according to their preferences. This trend is also seen in tangible privacy research that provides evidence that users with a high affinity for technology interaction prefer tangible mechanisms~\cite{delgado2024you}.\\
    \item \includegraphics[height=20pt,width=20pt]{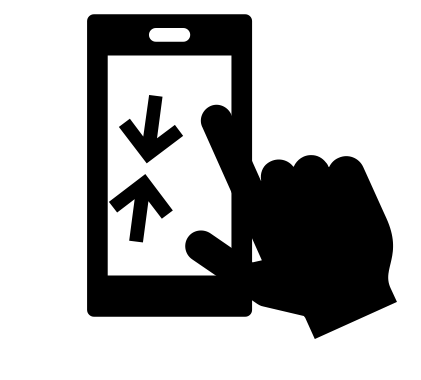} \textbf{Importance of the Out-of-Device Privacy: }Users with high ODPS scores preferred usefulness and personalisation of privacy protection and vice versa. This shows that people who are highly concerned about their out-of-device privacy would need privacy protection more than people with low importance for out-of-device privacy~\cite{farzand2024out}. This finding resonates with profiling users based on their privacy profiles~\cite{marky2024decide}. \\
    \item \includegraphics[height=20pt,width=20pt]{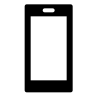} \textbf{Designing Discrete Mechanisms: }Among all mechanism categories presented to participants, participants favoured discreet mechanisms the most especially icon overlay followed by haptic and tangible. This aligns with previous work that reported that users prefer unobtrusive and alerting mechanisms~\cite{farzand2021interplay,farzand2022shoulder}.

\end{itemize}

\subsection{Future Work}
In this paper, we explored how personal attributes such as gender, affinity for technology, or importance for out-of-device threats impact user preferences for protection mechanisms against shoulder surfing. The next step in this direction invites researchers to validate the relationship findings by running studies in the wild. Running studies in the wild would also help to overcome the potential presence of privacy paradox in users' responses. Further, the results of our study showed that overall, participants mainly preferred icon overlay, haptic, and tangible mechanisms. Considering users' preferences, another future research direction is to explore how the current design of icon overlay, haptic, and tangible can be improved to offer an improved privacy experience. Some interesting research directions to explore include what information can be conveyed to the user about the bystander through these mechanisms and how these mechanisms impact user experience. For example, in the case of haptics, the vibration pattern has to be different from the standard vibration pattern so that the user can distinguish between general notifications and bystander alert notifications. This becomes more complicated when users have customised vibrations for different notification types. In such cases, it can be hard for the user to keep track and remain aware of which vibration format conveys what information. In the case of tangible mechanisms, this mechanism category can be further expanded by including phone accessories (for example, phone covers) as bystander alert mechanisms. Prior work has investigated using tangible items for authentication and shown promising results~\cite{marky20203d}; exploring phone accessories as tangible mechanisms for communicating with bystanders might reveal interesting results as well.

\section{Conclusion}
In this paper, we aimed to explore correlations between personal attributes and preferences for protection mechanisms against content-based shoulder surfing. For this, we first identified existing protection mechanisms through a systematic literature review. We then categorised them based on design similarities. We then presented the mechanism categories to 192 participants and inquired about their preferences. Our results showed that users agree that the presented mechanisms assist in protecting their privacy but they would prefer using non-digital alternatives such as using the hand to cover the screen. Moreover, among the mechanisms, participants mainly preferred icon overlay. We also found that no significant differences exist in preferences between male and female iOS and Android users. Based on the results, we present design guidelines to support the design of future protection mechanisms.

\bibliographystyle{ieeetr}
\bibliography{references}

\end{document}